# Modeling water supply networks and gastrointestinal disorder symptoms with CAR models

# HUNT Study, Norway

**Håvard Wahl Kongsgård 1**
havard@beekeeper.no

**Geir-Arne Fuglstad 2**
fuglstad@math.ntnu.no

**Håvard Rue 3**
Haavard.Rue@kaust.edu.sa

**Kristian Hveem 1**
kristian.hveem@ntnu.no

**Steinar Krokstad 1**
steinar.krokstad@ntnu.no

**1 Department of Public Health and General Practice, NTNU**
**2 Department of Mathematical Sciences, NTNU**
**3 KAUST - King Abdullah University of Science and Technology**

**Abstract**

*Background:*
The direct modeling of water networks is not a common practice in modern epidemiology. While space often serves as a proxy, it can be problematic. There are multiple ways to directly model water networks, but these methods are not straightforward and can be difficult to implement. This study suggests a simple approach for modeling water networks and diseases, and applies this method to a dataset of self-reported gastrointestinal conditions from a questionnaire-based population health survey in central Norway.

*Method:*
Our approach is based on a standard conditional autoregressive (CAR) model. An inverse matrix was constructed, with nodes weighted based on the distance to neighboring nodes within the networks. This matrix was then fitted as a generic model. To illustrate its possible use, we utilized data taken from a questionnaire-based population health survey, the HUNT Study, to measure self-reported gastrointestinal complaints. For hypothesis testing, we used the deviance information criterion (DIC) and included variables in a stepwise manner.

*Results:*
The full model converged after six hours. We found no relation between the water networks and the health conditions of people whose residences connected to different parts of the network in the geographical area studied.

*Conclusion:*
All water network models are simplifications of the real networks. Nevertheless, we suggest a valid approach for distinguishing between the general spatial effect and the water network using a generic model.

*Keywords:* Water network, graph, spatial effect, disease mapping.

**Introduction**

Access to clean and safe water is essential to sustaining a healthy life. The quality of tap water is generally excellent in Northern Europe. However, in Norway in particular, a colder climate combined with an aging pipe system can lead to cracked and damaged pipes. The small leaks this damage causes can potentially allow the contamination of tap water by sewage or soil (Myrstad et al., 2014). Previous research has found that a fall in pressure due to maintenance and leaks is a health hazard (Nygård et al., 2007). A global meta-study of the same problem found that water outages are also associated with increased gastrointestinal illness, suggesting a potential health risk for consumers served by piped water networks (Ercumen et al., 2014). However, because the effects of these leaks are often so small and take place over a long period, it is often difficult to document such health-related incidents.

To conduct an analysis of a water supply, the water system is broken down into three components: source, treatment, and delivery. While water source and treatment are single-point events, the delivery of water poses a more complex, network-oriented problem. Given this complexity and the dependency of nodes that share information, these networks are notoriously difficult to analyze. To understand and explore such networks, researchers often use proxies (Gog et al., 2014).

For example, in 1855 Dr. John Snow used spatial data to understand cholera outbreaks and their causes. He discovered that the actual causal mechanism was a contaminated water pump shared by hundreds of people, yet the effect was strong enough to have a spatial signature. However, in statistical models, the effects of individual health factors are often tiny, making it very hard to distinguish the signal from the noise. Additionally, using space as a proxy can be problematic because information is lost in the process. The exploration of modern travel patterns and the spread of pandemics is a useful tool for understanding this problem—although diseases spread quickly between continents, the data may not reveal any spatial patterns. Therefore, a proxy cannot be used to understand how a disease spreads and a model of the network should be used instead (Gog et al., 2014).

**Aims**

The aim of this study is to illustrate how a model based on the conditional autoregressive (CAR) approach can be used to analyze water supply pipeline networks. As a test, we applied this model to data from the Nord-Trøndelag Health Study (HUNT Study).

**Methods**

Using the location of pipeline segments, it is possible to create an artificial representation of the water network by creating a new set of coordinates or a distance matrix that represents the real relationship between nodes. This approach is very similar to Euclidean distance (the ordinary distance between two points, for example; that is, a straight line) in contrast with driving distance. Although theoretically we could create a nonlinear coordinate system, the practicalities of designing such a system are unclear. Additionally, the approach and solution would differ from case to case.

An alternative is to observe node states and hidden flows in the network. Many physics methods could be applied to water networks in general (Wang et al., 2017); however, this approach does not conventionally allow additional factors such as the age or gender of the consumers to be taken into account, making it unsuitable for use in epidemiology.

As an alternative solution, we based our approach on a standard CAR model (Rue & Held, 2005). As illustrated in Figure 1, a CAR model is typically used to capture the spatial effect of bordering provincial units, which, for this study, is the dependence between observations.

*Construction of the inverse matrix*

In a conventional CAR model, units whose edges touch are considered spatially linked and dependent. This relationship is used to construct an inverse matrix, and the matrix is fitted with the rest of the data to account for and measure the dependency between observations. In constructing the inverse matrix, bordering relationships are given the value -1, and each unit is made up of the positive sum of the number of bordering units. Thus, in our experiment, Province A has two neighbors, B and C (Figure 1). The relationship between B or C and A is therefore -1, and A equals 2 (Table 1).

Water networks are directed graphs. In the case of contamination, only nodes along the directional flow are affected (Figure 2). With modeled water networks, it is possible to use a similar approach to a conventional CAR model. However, as nodes are seldom evenly distributed within the network, it is better to divide -1 by the distance between the nodes and then calculate the sum. If we consider the network in Figure 3, the values for the nodes would be A = 0.13 (Table 2).

*Hypothesis testing*

The hypothesis testing of random nonlinear effects is not always straightforward. Depending on how the model is fitted, it is possible to use conventional methods such as P-values or odds ratios. As an alternative, we could also use the Akaike information criterion (AIC), Bayesian information criterion (BIC), or deviance information criterion (DIC) (Anderson & Burnham, 2006). A lower value in each of these implies a better model. A reduction value of 10 or more is the most common significant threshold. Only models based on the same data and observations can be compared, as it is not possible to make comparisons across different datasets (Bolker et al., 2009).

*Data*

In this study, we used data from the HUNT Study (HUNT3, 2006-2008), a regional health study providing a unique database of medical histories (Krokstad, 2013). While the study area was predominantly rural, it included five additional municipalities with towns of up to 21,000 inhabitants (Krokstad, 2013). Using this dataset, we extracted data from three bordering municipalities: Levanger, Verdal, and Steinkjer. We obtained detailed maps of the water pipeline networks from the municipalities and constructed a new graph using these schematics, removing all unnecessary nodes from the original network. We then constructed a sparse inverse matrix as described above, adding numbers that linked the nodes in the network to the participants.

The HUNT3 Study data contained multiple self-reported medical conditions and their severity. For example, we extracted data on diarrhea and reflux, two common gastrointestinal conditions that may be linked to the drinking of contaminated water. For the sake of simplicity and to normalize the statistical distribution, we disregarded the severity and recoded reported cases of diarrhea and reflux as single binary events.

*Models*

To fit the model, we used the R-package R-INLA, which includes built-in support to handle sparse matrixes (Rue et al., 2009), including various spatial approaches and other nonlinear effects. We adjusted for gender and age in the model. To distinguish between the spatial structure of the pipelines and the general spatial effect, we included a nonlinear effect for space. To account for social and genetic factors, a variable for people living in the same household was also included. We added variables to the model in a stepwise manner to observe the change in DIC.

In terms of the resources needed for a trained researcher to run the statistics in practice, fitting a basic model to age, sex, and water network data (Table 2) took 16 minutes, and the full model took six hours to complete.

**Results**

Table 3 shows the six different models that were fitted to the data. As the table shows, starting with the baseline model with age and gender (DIC 20802/15333), the DIC value increased as we added additional effects, including the water graph (DIC 20886/15572 for age, gender, and water graph). This implies that the predictive power of the model did not improve and that no information passed between participants via the network. Therefore, no link between the reported conditions and the water network was observed.

**Discussion**

As expected, we found no relation between the water networks and gastrointestinal health conditions. For this study, we used a CAR-like model to analyze variations in reported diarrhea and reflux symptoms along different segments of the water supply pipeline networks. Our data showed that the conditions occurred randomly within the water network and space. The study provides a useful demonstration of how water networks can be analyzed; however, the usefulness of this approach depends heavily on the data.

Generally, the ideal scenario would be a situation in which one observes a spatial effect and seeks to investigate the pattern further—a common practice used to investigate the origin of outbreaks in real-life scenarios (Braeye et al., 2017). In this case, the dataset provided was not ideal as it came from a general cohort and the conditions were self-reported. Moreover, our representation was a simplification; it did not capture all of the properties of the original. In this situation, it is difficult to detect faint signals as the water networks are complex and the correlation between environmental factors and health conditions is weak. Therefore, in practice, even if there were a tiny effect, the signal would be lost in the noise. Because of this, we failed to detect a spatial effect for both conditions (diarrhea and reflux).

Another limitation in the study was the focus on measuring the global effect of a contaminated water supply. From a public health point of view, local effects can sometimes be a more revealing subject for research than global effects. Measuring the effect of areas and segments of the water network where conditions cluster is often more informative than analyzing that of the entire network. While it is possible to search for outliers and nodes with strong effects, it is better to construct and then fit a similar matrix using common cluster algorithms or related methods (Pedregosa et al., 2011).

Nevertheless, in a real-life scenario, a single model will never give a definitive answer since cluster and spatial effects can have multiple origins.

**Conclusion**

Diseases transmitted through water networks pose a complex problem. Our approach is one of many tools of potential value to researchers. In essence, it allows researchers to distinguish between the general spatial effect and the effect of networks. It can be useful for datasets with unexplained spatial effects in order to rule out, or confirm, possible links between networks generally and diseases specifically. Nevertheless, a single model is insufficient and more investigations are necessary. Likewise, this approach mainly concerns global effects, and it is important to recognize that local effects can often be equally, if not more, significant.

**Figures and tables:**

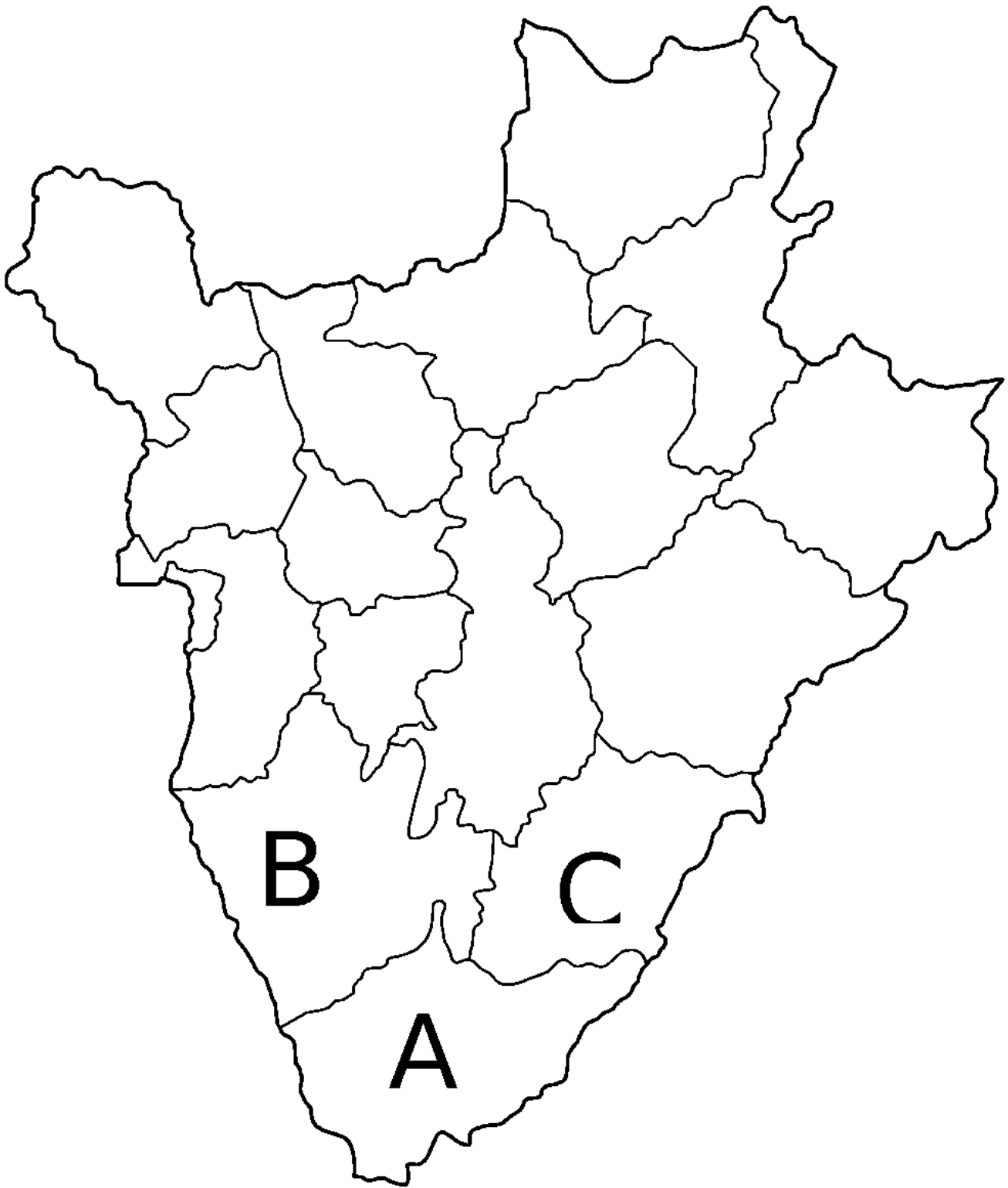

*Figure 1: Example of a CAR model. This uses provinces A, B and C.*

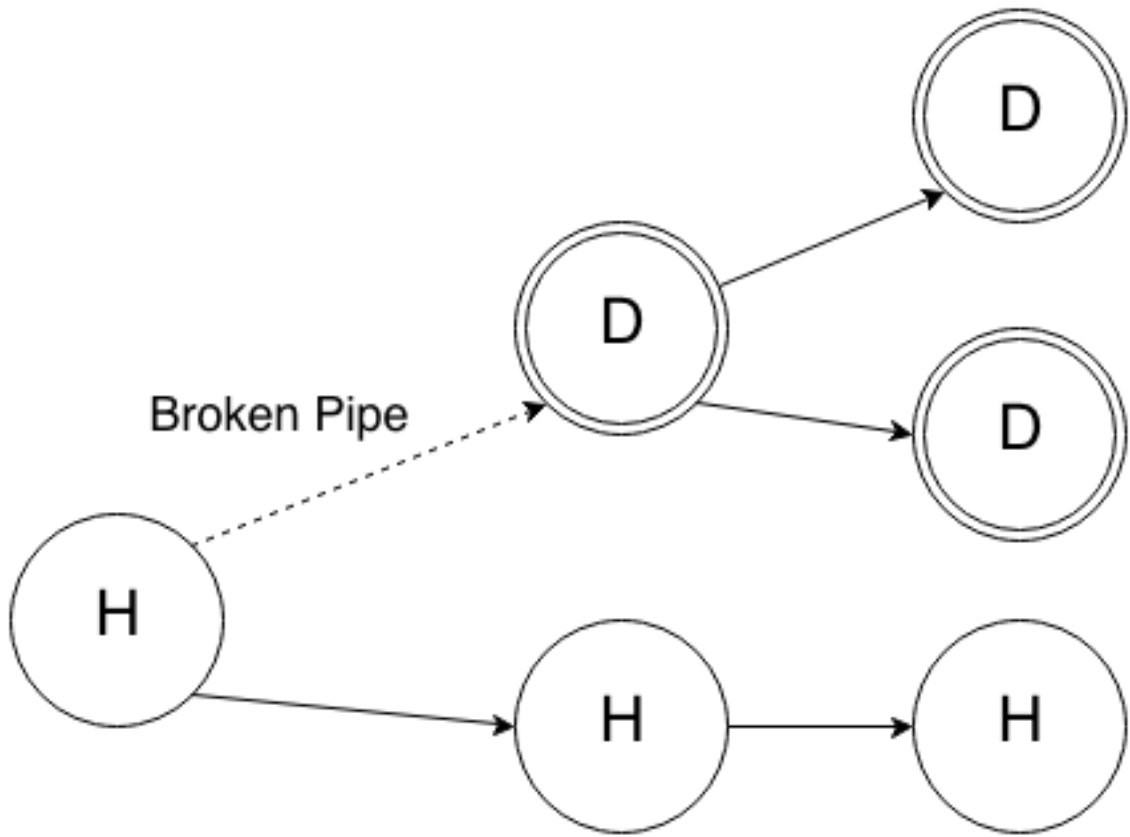

*Figure 2: The effect of cracked pipes in a water network. Healthy nodes (H) and nodes with potentially contaminated water (D) are identified.*

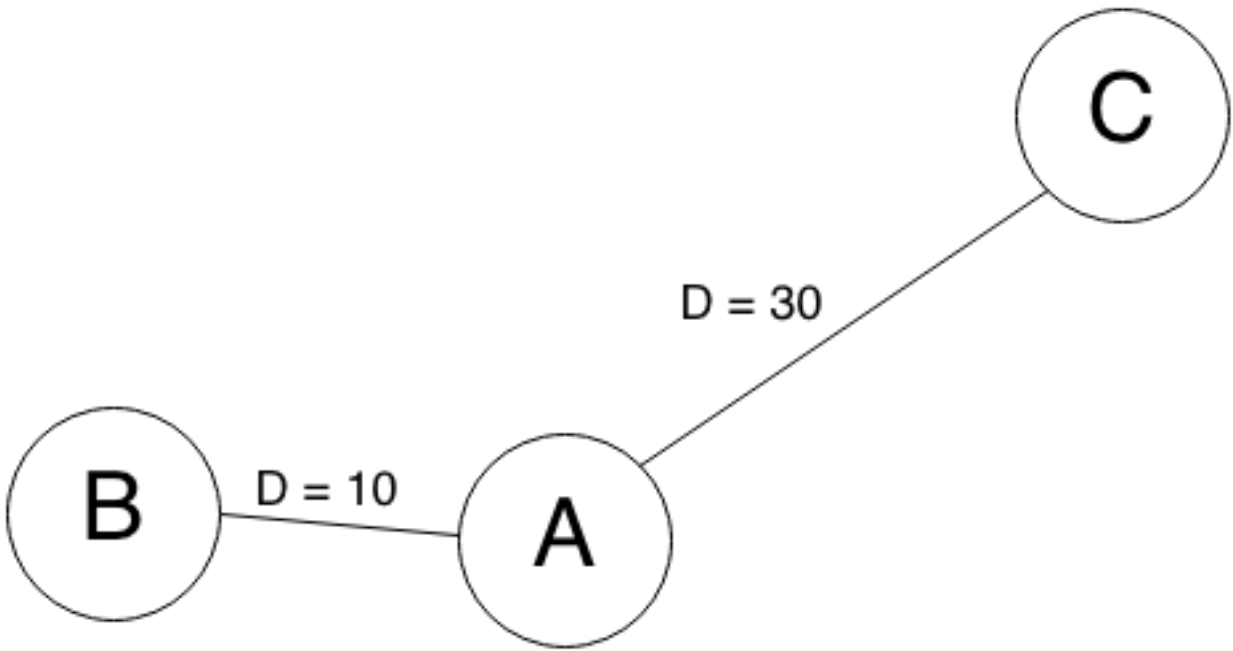

*Figure 3: Simple network with nodes A, B, and C. Distances between edges are also indicated (D).*

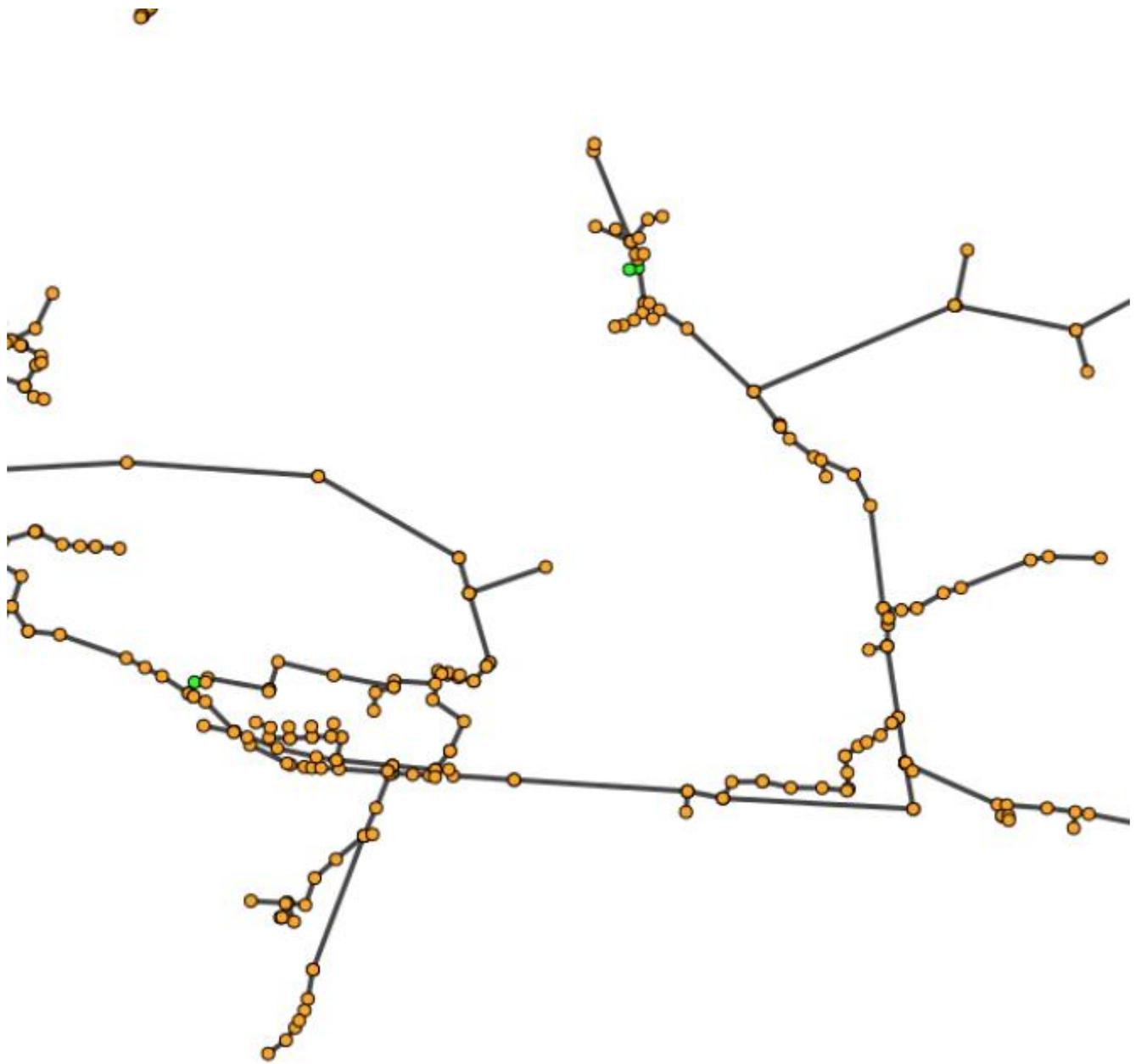

*Figure 4: Simplified water network with nodes and edges.*

**Table 1: Inverse matrix for Region A, B, & C for map in Figure 1.**

|   | A  | B  | C  |
|---|----|----|----|
| A | 2  | -1 | -1 |
| B | -1 | 5  | -1 |
| C | -1 | -1 | 4  |

**Table 2: Inverse matrix for the network in Figure 3.**

|   | A     | B    | C     |
|---|-------|------|-------|
| A | 0.13  | -0.1 | -0.03 |
| B | -0.1  | 0.1  | 0     |
| C | -0.03 | 0    | 0.03  |

**Table 3: Model comparison via DIC value; lower is better**

| Model | Diarrhea | Reflux |
|---|---|---|
| *Age, Gender* | 20,802 | 15,333 |
| *Age, Gender, House ID* | 20,803 | 15,333 |
| *Age, Gender, House ID, Spatial Effect* | 20,808 | 15,331 |
| *Age, Gender, House ID, Spatial Effect, Water Graph* | 20,891 | 15,382 |
| *Age, Gender, Water Graph* | 20,886 | 15,386 |
| *Water Graph* | 20,904 | 15,572 |
| *N:* | *15,281* | *11,431* |